\documentclass[11pt]{article}
\usepackage{moriond,epsfig}

\def\ra{\rightarrow}

\def\be{\begin{equation}}
\def\ee{\end{equation}}
\def\bea{\begin{eqnarray}}
\def\eea{\end{eqnarray}}

\begin{document}

\newcommand{\de}{\Delta E}
\newcommand{\mbc}{M_{\rm bc}}
\newcommand{\bb}{B{\bar B}}
\newcommand{\qq}{q{\bar q}}

\vspace*{4cm}
\title{\Large New resonances in B-meson decays}

\author{ R. CHISTOV \\ (for the Belle Collaboration)}

\address{Institute for Theoretical and Experimental Physics, Moscow, Russia}

\maketitle\abstracts{
     The $X(3872)$ and $Y(3940)$ properties and decay modes from Belle are reviewed  
and the results on a search for the decay $B^+\ra h_c K^+$, $h_c\ra \eta_c~\gamma$ 
at Belle are presented. 
}

\section{Introduction}

$B$-mesons have proved to be a rich source of new particles. The B-factories
 at KEK (Belle) and SLAC (BaBar) extensively 
use the exclusive production of $J/\psi$, $\chi_{c1}$ and $\eta_c$ charmonia for 
the CP violation measurements. 
In addition to these conventional charmonia Belle observed 
the production of $\eta_c(2S)$~\cite{olsen1} and $\psi(3770)$~\cite{me} in exclusive $B\ra (c\bar{c})K^+$ decays. 
Belle and BaBar also observed inclusive $\chi_{c2}$ production in 
$B$ decays~\cite{chi_c2}.   
In 2003, by analyzing the $B^+\ra J/\psi~\pi^+~\pi^-~K^+$ decays, 
Belle observed a narrow charmonium-like new state 
(denoted as $X(3872)$) decaying into $J/\psi~\pi^+\pi^-$~\cite{belle_x3872_1}. 
Recently Belle reported the observation of $Y(3940)\ra J/\psi~\omega$ 
in $B^+\ra Y(3940)~K^+$ decays~\cite{y3940}.  

All observed resonances -- from $\eta_{c}(2S)$ to $Y(3940)$ -- 
are difficult to reconstruct without the constraints
provided by $B$ decays, as they decay 
to high-multiplicity final states. 
$B$ mesons also provide an excellent 
opportunity to test different hypotheses for the $J^P$ quantum 
numbers of 
these resonances via decay angle analysis.  

\section{Observation of  $X(3872)$ and it's properties}

Just after the discovery of $X(3872)$ by Belle, this new state was confirmed by the CDF, 
D0 and BaBar collaborations~\cite{cdf-d0}. 
Its mass was measured to be $3871.9\pm 0.5$~MeV which is very close to the $D^0\bar D^{*0}$ 
threshold of $3871.3\pm 1.0$~MeV.
All these first measurements directly provide a lot of information on the 
$X(3872)$ properties~\cite{belle_x3872_1}:
\begin{itemize}
\item the two pion mass from the $X(3872)\ra J/\psi~\pi^+~\pi^-$ decay tends 
to peak near the $\rho^0$ mass, consistent with positive C-parity of the $X(3872)$;
\item the decays $X(3872)\ra\chi_{c1,2}~\gamma$ are not seen; this 
likely excludes the $1^3D_{2,3}$ ($\psi_{2,3}$) assignment for the $X(3872)$;
\item the decay $X(3872)\ra D\bar{D}$ is suppressed or forbidden~\cite{me}; this, together with its narrow width ($\Gamma<2.7$~MeV), 
suggests $J^P=0^+, 1^-, 2^+, ...$ are ruled out;
\end{itemize}

CDF and D0 have measured the $X(3872)$ production properties 
to be very similar to those of the $\psi(2S)$~\cite{cdf-d0}.
BaBar reported a null search for the charged $X(3872)$ partners 
in $B\ra J/\psi~\pi^+~\pi^0~K^+$ decays~\cite{babar_charged_x_partners} 
and a null search for $X(3872)\ra J/\psi~\eta$ decay~\cite{babar_jpsieta}. 
The former rules out the isovector hypothesis 
and the latter excludes the presence of gluonic degrees of freedom in the $X(3872)$ wave function.

\begin{figure}
  \begin{center}
    \includegraphics[width=0.45\textwidth]{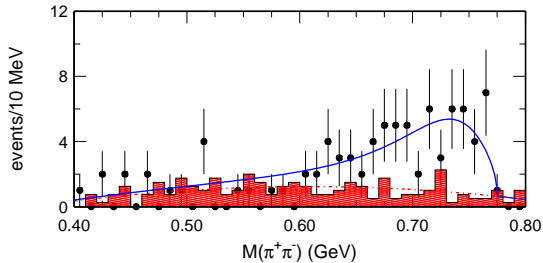}
    \caption{
The invariant mass spectrum of $\pi^+~\pi^-$ pairs from the 
$X(3972)\ra J/\psi~\pi^+~\pi^-$ decay. 
    }
    \label{belle_x_update}
  \end{center}
\end{figure}

Fig.~\ref{belle_x_update} shows the $M(\pi^+~\pi^-)$ 
spectrum from the updated analysis of $X(3872)\ra J/\psi~\pi^+~\pi^-$ decays by Belle (253~fb$^{-1}$).
The $\rho^0$ signal is strong and supports $C(X(3872))=+1$.  

Recently Belle has found evidence for another decay mode of $X(3872)$: 
$X(3872)\ra J/\psi~\omega^*$ where $\omega^*$ is virtual and reconstructed in the $\pi^+~\pi^-~\pi^0$ channel. 
According to Swanson's model~\cite{swanson}, this observation supports the 
$D^0\bar{D}^{*0}$ molecular interpretation of $X(3872)$.   
The full set of the most recent Belle results on the $X(3872)$ properties 
can be found elsewhere~\cite{olsen_new_x3872}. 
   
\section{Observation of $Y(3940)$}

\begin{figure}
  \begin{center}
    \includegraphics[width=0.75\textwidth]{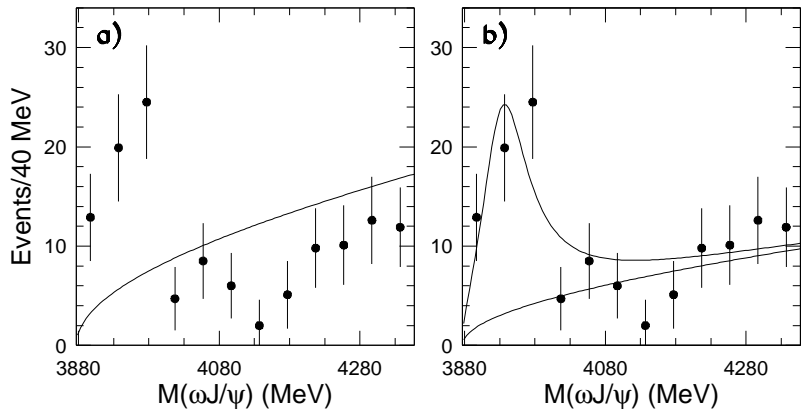}
    \caption{The invariant mass spectrum of $J/\psi~\omega$ combinations from the $B^+\ra J/\psi~\omega~K^+$ decay. The curve in (a) indicates the result of a fit with only a phase-space-like two-body threshold function. The curve in (b) shows the result of a fit that includes an S-wave Breit-Wigner resonance term.
    }
    \label{belle_y}
  \end{center}
\end{figure}

By analyzing exclusive $B^+\ra J/\psi~\pi^+~\pi^-~\pi^0~K^+$ decays 
Belle observed a new resonance $Y(3940)$  
decaying to $J/\psi~\omega$~\cite{y3940}. Fig.~\ref{belle_y} shows 
the $M(J/\psi~\omega)$ distribution for the $B$-meson candidates.  
The curve in Fig.~\ref{belle_y}~(a) indicates the result of a fit 
with only a phase-space-like two-body threshold function.  
The curve in Fig.~\ref{belle_y}~(b) shows the result of a fit that includes an S-wave Breit-Wigner resonance term.
The mass and width of $Y(3940)$ were measured to be 
$3943\pm 11\pm 13$~MeV and $87\pm 22\pm 26$~MeV, respectively. 
The observed state is above the $D\bar{D}^{(*)}$ threshold and would decay predominantly 
to $D\bar{D}$ and/or $D\bar{D}^{*}$ if it is a $c\bar{c}$ charmonium.  
In contrast, for a $c\bar{c}-gluon$ hybrid 
the open charm decay modes are suppressed or forbidden. 
So the observed $Y(3940)$ is a possible candidate for the first  
$c\bar{c}-gluon$ hybrid state. 


\section{Search for $B^+\ra h_c K^+$}

The $1^1P_1$, $J^P=1^{+-}$ $h_c$ has for a long time been a missing state. 
Recently CLEO reported the observation of $h_c$ in $\psi(2S)\ra h_c~\pi^0$, 
$h_c\ra\eta_c~\gamma$ decays~\cite{cleo_hc}.
The mass was measured to be $3524.4\pm 0.6\pm 0.4$~MeV - in agreement with 
theoretical expectations that the $M(h_c)$ is close to the c.o.g. of of the $<1^3P_J>$ triplet.

\begin{figure}
  \begin{center}
    \includegraphics[width=0.75\textwidth]{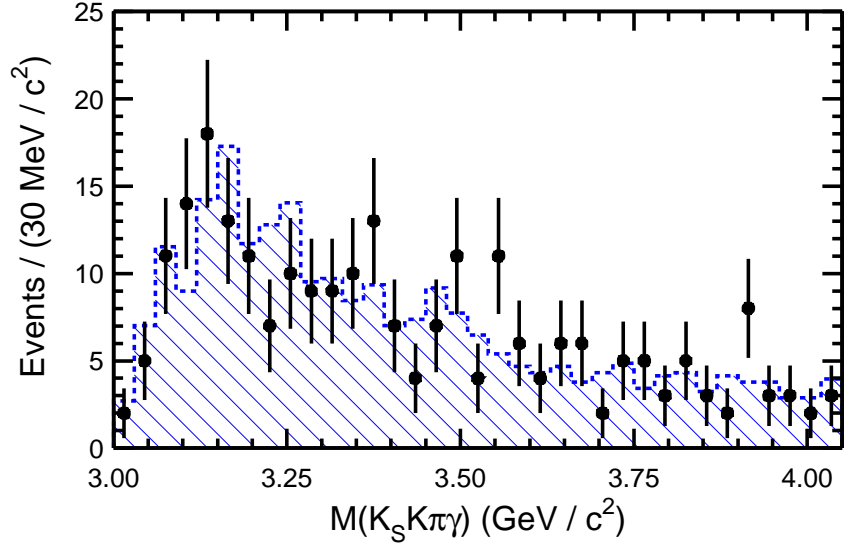}
    \caption{ The invariant mass spectrum of $\eta_c~\gamma$ for the 
$B^+\ra\eta_c~\gamma~K^+$ candidates.
    }
    \label{fangfang}
  \end{center}
\end{figure}

Belle searched for $h_c$ production in exclusive $B^+\ra h_c~K^+$, $h_c\ra\eta_c~\gamma$ decays. 
Fig.~\ref{fangfang} shows the $M(\eta_c~\gamma)$
for the $B$-candidates. No evidence for a signal around the CLEO $h_c$ mass is seen. 
Belle set an upper limit for 
${\cal B}(B^+\ra h_c~K^+)\times{\cal B}(h_c\ra\eta_c~\gamma)$ 
that is less than $1.5\times 10^{-4}$ for $M(h_c)\sim 3520$~MeV~\cite{belle_hc}.

\section{Study of $D_{sJ}(2317)$ and $D_{sJ}(2460)$}

The $D_{sJ}(2317)^{*+}$ and $D_{sJ}(2460)^{+}$ mesons were observed by 
BaBar~\cite{babar_dsj} and CLEO~\cite{cleo_dsj}. 
Belle confirmed these states and observed their production in 
exclusive $B\ra D^{*+}_{sJ}\bar{D}^{(*)}$ decays~\cite{krokovny1}. 
The observation of these decays allowed us to 
perform decay angle analysis. Belle data support 
 $J=0$ for  $D_{sJ}(2317)$ and $J=1$ for $D_{sJ}(2460)$~\cite{krokovny2}.

\section{Summary}

$B$ mesons provide a clean environment for the observation 
of yet-unseen charmonia and other new unexpected resonances and   
the understanding of their properties.

The nature of new resonances $X(3872)$ and  $Y(3940)$ remains unclear so far. They could be either $c\bar{c}$ states or exotic hadrons: $D^0\bar{D}^{*0}$ molecular ($X(3872)$) and $c\bar{c}-gluon$ hybrid ($Y(3940)$).

\section*{Acknowledgments}
We thank the KEKB group for the excellent operation of the
accelerator, the KEK cryogenics group for the efficient
operation of the solenoid, and the KEK computer group and
the NII for valuable computing and Super-SINET network
support.  We acknowledge support from MEXT and JSPS (Japan);
ARC and DEST (Australia); NSFC (contract No.~10175071,
China); DST (India); the BK21 program of MOEHRD and the CHEP
SRC program of KOSEF (Korea); KBN (contract No.~2P03B 01324,
Poland); MIST (Russia); MHEST (Slovenia);  SNSF (Switzerland); NSC and MOE
(Taiwan); and DOE (USA).

\end{document}